\documentclass[conference]{IEEEtran}
\usepackage{graphicx}
\usepackage{url}
\usepackage{amsmath, amsfonts}
\usepackage{amsthm, amssymb}
\usepackage{comment}
\usepackage{xcolor}
\usepackage[font=small]{caption}
\usepackage{subcaption}
\usepackage{array}
\usepackage{dsfont}
\usepackage{cite}

\DeclareMathAlphabet\mathbfcal{OMS}{cmsy}{b}{n}

\ifCLASSINFOpdf
\else
\fi
\newcommand\blfootnote[1]{%
  \begingroup
  \renewcommand\thefootnote{}\footnote{#1}%
  \addtocounter{footnote}{-1}%
  \endgroup
}

\begin{document}


\title{Evaluating Vulnerabilities of Connected Vehicles Under Cyber Attacks by Attack-Defense Tree}

\author{\IEEEauthorblockN{Muhammad Baqer Mollah$^1$, Honggang Wang$^2$, and Hua Fang$^3$}
\IEEEauthorblockA{$^1$Dept. of Electrical and Computer Engineering, University of Massachusetts Dartmouth, MA 02747 \\ $^2$Dept. of Graduate Computer Science and Engineering, Yeshiva University, NY 10016 \\
$^3$Dept. of Computer and Information Science, University of Massachusetts Dartmouth, MA 02747 \\
Emails: mmollah@umassd.edu, honggang.wang@yu.edu, and hfang2@umassd.edu}
        }

\markboth{}%
{Shell \MakeLowercase{\textit{et al.}}: Bare Demo of IEEEtran.cls for IEEE Journals}

\maketitle

\begin{abstract}
    Connected vehicles represent a key enabler of intelligent transportation systems, where vehicles are equipped with advanced communication, sensing, and computing technologies to interact not only with one another but also with surrounding infrastructures and the environment. Through continuous data exchange, such vehicles are capable of enhancing road safety, improving traffic efficiency, and ensuring more reliable mobility services. Further, when these capabilities are integrated with advanced automation technologies, the concept essentially evolves into connected and autonomous vehicles (CAVs). While connected vehicles primarily focus on seamless information sharing, autonomous vehicles are mainly dependent on advanced perception, decision-making, and control mechanisms to operate with minimal or without human intervention. However, as a result of connectivity, an adversary with malicious intentions might be able to compromise successfully by breaching the system components of CAVs. In this paper, we present an attack-tree based methodology for evaluating cyber security vulnerabilities in CAVs. In particular, we utilize the attack-defense tree formulation to systematically assess attack-leaf vulnerabilities, and before analyzing the vulnerability indices, we also define a measure of vulnerabilities, which is based on existing cyber security threats and corresponding defensive countermeasures.  \blfootnote{This paper has been accepted to present at International Conference on Computing, Networking and Communication (ICNC), Maui, Hawaii, USA, 2026.}
\end{abstract}


\IEEEpeerreviewmaketitle

\section{Introduction}
    \blfootnote{\copyright 2026 IEEE. Personal use of this material is permitted. Permission from IEEE must be obtained for all other uses, in any current or future media, including reprinting/republishing this material for advertising or promotional purposes, creating new collective works, for resale or redistribution to servers or lists, or reuse of any copyrighted component of this work in other works.} Connected and autonomous vehicles (CAVs) are expected to reach a new level of capability with the advancement of emerging technologies, primarily shaped by cyber-physical systems, which integrate computational and physical components, are interconnected through internal in-vehicle networks as well as external networks to enable real-world sensing, control, and the provision of interconnected applications and services \cite{mollah2024mmwave}. The convergence of connectivity and autonomy develops a powerful synergy, where connectivity enhances situational awareness by providing access to information beyond the line of sight, whereas autonomy ensures real-time decision-making and actuation based on both local sensing and shared data. However, such a new settings involving increased connectivity, digitization, and complex platforms, such as hardware and software might introduce cyber threats \cite{darwaish2025security} and new attack vectors at the same time \cite{10643680}, which demand strict cyber security concerns to be addressed. With this in mind, insufficient monitoring and measures could not only put the assets of connected vehicles at risk, but also endanger human life and have potential severe consequences for people's safety.
    
    Essentially, vulnerability evaluation is necessary for CAVs due to their dependencies on complex cyber-physical systems. Since these vehicles are required to interact continuously with internal networks, external infrastructures, and other vehicles, a wide spectrum of cyber threats might be exposed. By systematically identifying and evaluating vulnerabilities, it is possible to quantify the security risks, prioritize countermeasures, and design more resilient systems. Such assessments not only help in mitigating current threats but also provide insights for developing proactive defense mechanisms against evolving attack vectors, thereby ensuring the safety, reliability, and trustworthiness of future intelligent transportation systems.
    
    Attack-tree modeling \cite{konsta2024survey} provides an effective framework for systematically analyzing and evaluating the potential vulnerabilities within a system. The attack-trees are typically presented as a tree-like structure, where the root node and leaf nodes are represented the attacker's goal and the individual steps that should be taken to achieve that goal, respectively. And, the nodes in-between the root and leaf nodes indicate the intermediate steps that have been taken. By identifying all possible ways to attack a system, it is possible to focus the efforts to protect the most vulnerable areas. Also, the attack trees can be used to assess the likelihood and impact of different attacks, which can then be used to prioritize security measures and make informed decisions about how to protect a system.
    
    On the other hand, an attack-defense tree \cite{birchler2024attack,arias2024optimal} is also a graphical representation of all possible ways to attack a system, but it additionally includes the defenses as countermeasures for the purpose of preventing or mitigating those attacks, which can be further quantified using mathematical tools. Similar to an attack-tree, an attack-defense tree is essentially a valuable tool to identify and prioritize security risks and to design effective security measures. A number of recent studies have employed either attack tree or attack-defense tree modeling in various domains such as AI/ML based systems \cite{veria2024threat}, unmanned aerial vehicles \cite{sharma2024secure,kuzmin2023game,alexandre2023cybersecurity}, power systems/smart grid \cite{liu2025deep}, blockchain networks \cite{naik2022evaluation,eisentraut2021assessing}, and cyber-physical systems \cite{abdulhamid2025quantitative,teixeira2024model}, Internet of Things \cite{sacchetti2025attackdefense}. However, its application within the CAVs domain remains largely unexplored, which motivates us to investigate and extend this methodology for assessing vulnerabilities and security challenges in CAVs. To this end, this paper presents attack-defense modeling to systematically evaluate cyber security vulnerabilities in the context of CAVs. The key contributions can be summarized as follows.
    
\begin{itemize}
    \item A short literature review on recent cyber security and potential measure techniques for CAVs is presented.
    \item An attack-defense tree-based formulation is developed to identify and categorize potential attack paths and their corresponding countermeasures across individual attack-leaf vulnerabilities.
    \item Attack-defense modeling is particularly employed to assess the effectiveness of existing security measures and to explore improved countermeasures.
    \item A metric for quantifying vulnerabilities in CAVs is also defined, taking into account existing cyber security threats and corresponding defensive countermeasures.
    \item Finally, the impact of different attacks and defense strategies is also analyzed quantitatively, providing insights into the robustness of CAVs security architectures.
\end{itemize}

\section{System Model and Cyber Threats of CAVs}
    This section presents the system model and outlines the recent cyber threats relevant to connected and autonomous vehicles (CAVs) as follows.
    
\begin{figure} [!t]
	\includegraphics[width=\linewidth]{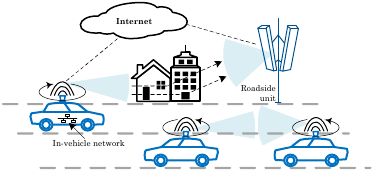}
	\centering
	\caption{System model components within the CAV ecosystem.}
    \label{fig: system}
\end{figure}
    
\subsection{System Model}
    In this work, we present the overall connected autonomous vehicle (CAV) ecosystem model along with its components, as illustrated in Fig. \ref{fig: system}. In particular, it consists of both fixed and mobile entities to support connected and autonomous driving. On the infrastructure side, a stationary base-station, which is commonly referred to as a roadside unit deployed along the roadway to provide communication services to the vehicles traveling within its coverage range. This RSU is equipped with storage devices and computational resources to support on processing vehicular data, run safety-critical applications, and provide vehicular services, such as HD map updating, cooperative perception, etc. On the vehicular side, each CAV carries a set of in-vehicle subsystems, including Electronic Control Units (ECUs), Controller Area Network (CAN) buses, sensors, and actuators. These subsystems together allow the vehicle to perceive, interpret, and interact with its environment. Furthermore, both RSUs and vehicles are equipped with V2X communication capabilities and a number of sensors, such as RGB cameras, LiDAR, GPS, and RADAR, which continuously generate and exchange multimodal information about the surrounding environment and other road participants.

    Based on this system model, this work presents a structured basis for analyzing the security overview of CAV operations and identifying how different threats propagate across interconnected components. Moreover, the combination of wireless communication, distributed sensing, and onboard control introduces multiple ways for adversaries to compromise the system components if proper countermeasures are not ensured. In specif, to analyze adversarial capabilities in in a structured way, we consider that all communications among CAV ecosystem entities take place over a public and potentially entrusted communication channel. And, in line with this assumption, our work adopts the Dolev–Yao threat model \cite{dolev1983security}, which grants adversaries full control over the communication medium. Under this model, attackers can passively monitor transmissions between entities or actively manipulate them by injecting forged packets, replaying previously captured messages, altering message contents, or impersonating legitimate nodes within the system. This well-established and widely accepted adversarial assumption basically allows us to assess how various vulnerabilities, attack methods, and defense mechanisms affect the overall security of the CAVs.

\subsection{Cyber Threats in CAVs}
    Considering the aforementioned system model, the CAV ecosystem might be exposed to a number of threats due to depending mainly on external connectivity with roadside units and other connected vehicles on the road. IN particular, such attacks generally can be categorized as (i) external vs. internal adversaries and (ii) passive vs. active adversaries \cite{alqahtani2024cybersecurity,10771587}. Here, the external adversaries are entities operating outside, such as remote attackers, rogue roadside units, or malicious vehicles, which might attempt to compromise the system without prior privileged access. In contrast, the internal adversaries already might have legitimate access or successfully exploited an in-vehicle subsystem that basically enable them to manipulate messages or system behaviors from inside. On the other hand, passive attacks involve silent observation of communication channels (e.g., eavesdropping, traffic analysis) without altering messages, whereas active attacks intentionally modify, inject, block, or fabricate messages to cause operational disruption.
    
    A significant category of internal and active threats involves in-vehicle intrusions, where adversaries compromise the internal communication architecture of a CAV. Most modern vehicle's in-vehicle networks rely on a wide of number of ECU, sensors, actuators, etc. and connected with the Controller Area Network (CAN) bus, which might lack typical security safeguards. Hence, the attackers who gain access through compromised in-vehicle subsystems, diagnostic ports, or software vulnerabilities might perform multiple manipulation strategies. Examples include (i) replay-based CAN compromises, where previously captured valid frames are retransmitted to mislead ECUs, (ii) excessive CAN frame injection, where adversaries flood the bus with high-priority messages to suppress legitimate commands, and (iii) spoofed actuator commands, where falsified braking, steering, or powertrain instructions are injected to force unsafe maneuvers. These in-vehicle network attacks directly threaten vehicle control, making them some of the most dangerous forms of adversarial behavior in the CAV domain.

    Apart from in-vehicle network attacks, attackers may target roadside units, which play an important role in vehicular services and vehicle-to-infrastructure (V2I) communication. A rogue or compromised roadside unit can disseminate falsified traffic alerts, incorrect perception data, or manipulated safety messages, influencing the behavior of all vehicles within its coverage area. Adversaries may also exploit operating system weaknesses, outdated firmware, or insecure software services running on roadside units to gain persistent control. Such attacks then can allow adversaries to alter cooperative awareness messages, suppress safety broadcasts, or distribute misleading traffic information, ultimately degrading reliability across the broader transportation network. In many cases, these threats originate externally, but once an roadside unit is compromised, its behavior would be the similar as an insider attacker.

    The expansion of vehicle-to-everything (V2X) communication interfaces might introduce additional attack surfaces that connect both cyber and physical domains. A primary concern involves the misuse of digital certificates and theft of credentials. In a typical CAV ecosystem, each vehicle and RSU relies on the Public Key Infrastructure (PKI) to verify the authenticity of transmitted messages and establish trust between entities. If an attacker gains unauthorized access to certificates or credentials, they can impersonate legitimate vehicles or RSUs, and eventually enable them to inject false information into the network. Beyond credential compromise, a more sophisticated adversaries may also target edge-based AI models deployed in vehicles or RSUs for cooperative perception. By poisoning training data, manipulating model updates, or tampering with inference results, adversaries can perform wrong predictions. Since V2X interfaces form the backbone of information exchange, these attacks might propagate across multiple vehicles and roadside units in the network and intensify their impacts on misleading vehicles into making unsafe decisions and disrupt trusted communications.


\begin{figure*} [t]
	\includegraphics[width=.95\linewidth]{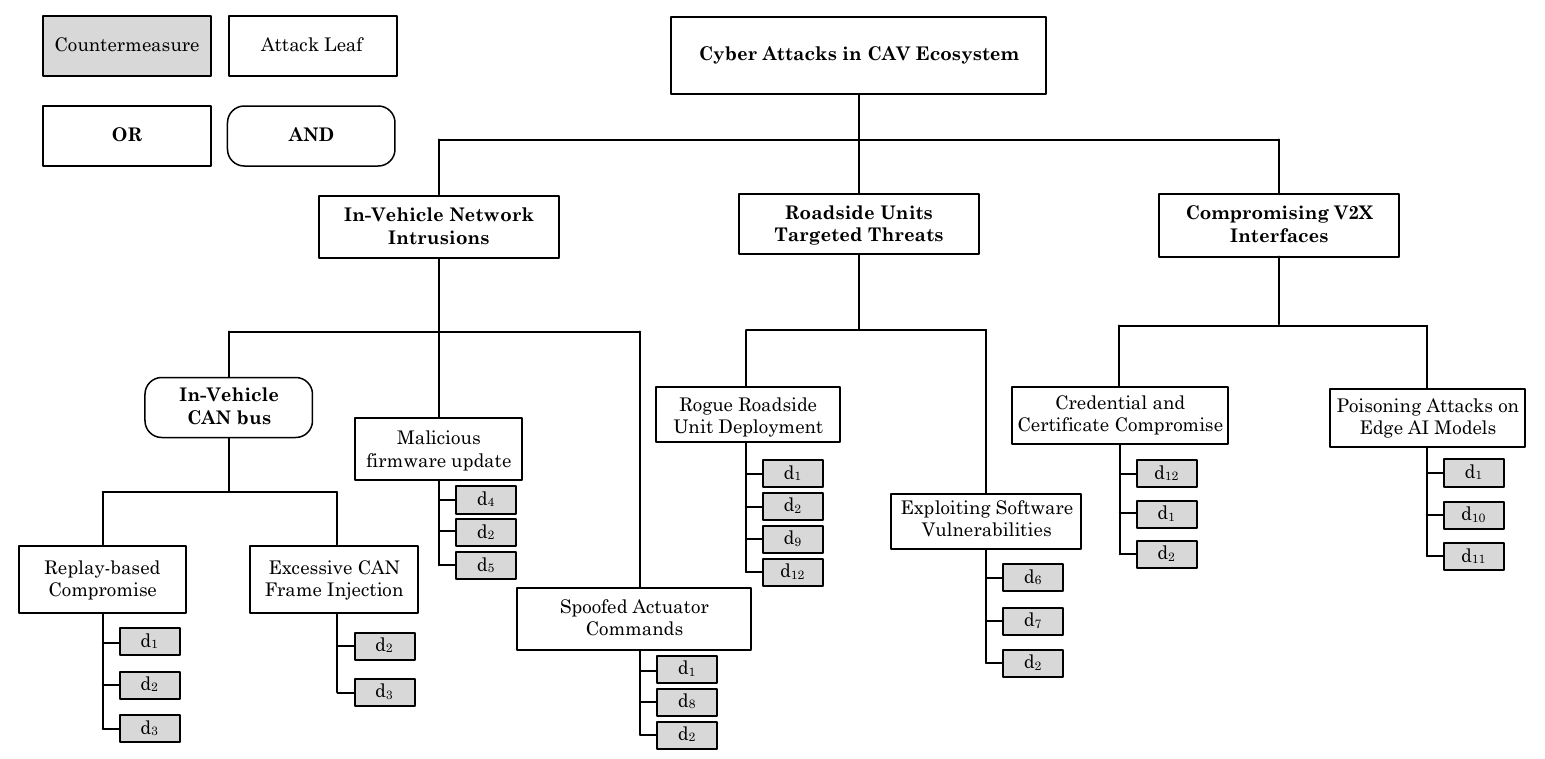}
	\centering
	\caption{The developed attack defense for CAVs with example attacks and defenses.}
    \label{fig: Attack-Defense}
\end{figure*}

\section{Attack-Defense Tree Modeling}
    This section introduces the attack-defense tree model and, based on it, presents a mathematical analytical framework for assessing vulnerabilities in CAVs.
    
\subsection{Attack-Defense Tree for CAVs}
    As previously mentioned, attack-defense trees are generally represented in a hierarchical, tree-like structure, where the root node denotes the attacker's ultimate goal, while the leaf nodes correspond to the specific actions required to achieve that goal. And, the intermediate nodes between the root and leaves represent the necessary steps or sub-goals along the attack path.

    In this work, we employ the cyber security modeling framework presented in \cite{ten2010cybersecurity}, which provides a structured methodology to represent and analyze cyber-physical vulnerabilities. The model basically facilitates the illustration of complex cyber-attack scenarios through hierarchical relationships between potential attack steps and their interdependencies. This representation helps in understanding not only individual attack paths but also the broader attack landscape that may target connected and autonomous vehicles (CAVs).

\begin{figure} [!t]
	\includegraphics[width=.90\linewidth]{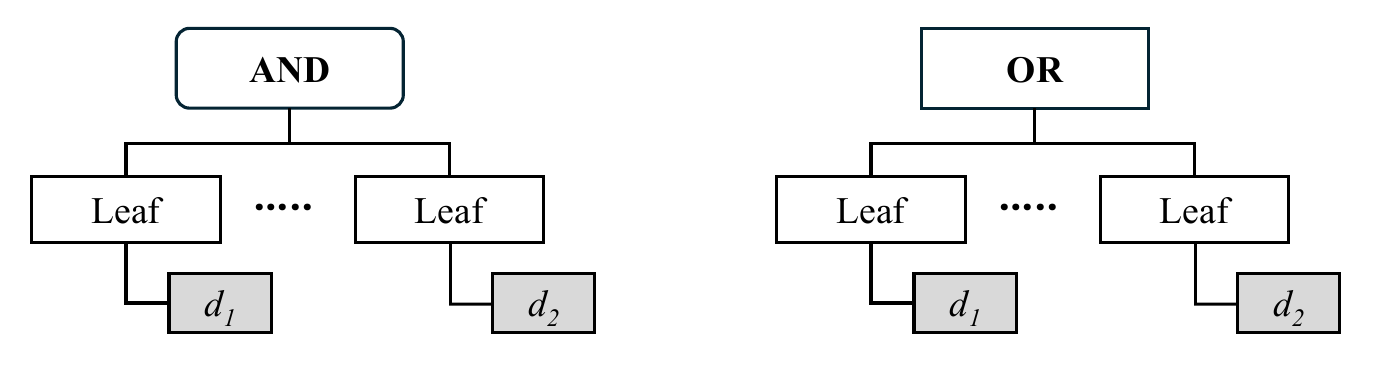}
	\centering
	\caption[The attack leaves are structured according to ``AND'' and ``OR'']{The attack leaves are structured according to ``AND'' and ``OR'' logical operators.}
    \label{fig: attack-tree2}
\end{figure}

    Fig. \ref{fig: attack-tree2} illustrates an example attack tree where the attack leaves are organized under logical “AND” and “OR” operators. Each leaf represents a specific attack step or vulnerability that contributes to the realization of a cyber threat. Furthermore, each leaf is associated with corresponding defensive countermeasures, denoted as $d_1$ and $d_2$. This logical structure enables a systematic assessment of potential attack paths and their corresponding defense mechanisms, helping to evaluate the effectiveness and interdependencies of different security strategies.

    Building upon the aforementioned model and the potential cyber threats discussed in the previous section, the developed attack-defense tree for CAVs is presented in Fig. \ref{fig: Attack-Defense}. This tree integrates both offensive and defensive elements, allowing for a comprehensive evaluation of cyber risks and mitigation approaches. In addition, Table \ref{tab: defenses} provides a summary of the defenses illustrated in Fig. \ref{fig: Attack-Defense}, where $\mathcal{D}$ denotes the complete set of defense mechanisms considered in this study. Eventually, these components offer a systematic way for analyzing the resilience of CAVs against a diverse range of cyber-attacks.

\begin{table}[!t]
    \footnotesize
    \centering
    \caption{Summary of defenses in the attack defense tree.}
    \label{tab: defenses}
    \begin{tabular}{m{2.5cm}|m{4.6cm}}
    
    \hline \hline

    \textbf{Countermeasures}, $\mathcal{D}$ & \textbf{Descriptions} \\
    \hline

    $d_1$    &    Cryptographic solutions \\
    \hline

    $d_2$    &    Intrusion detection systems \\
    \hline

    $d_3$    &    Access-control gateway \\
    \hline

    $d_4$    &    Secure boot and code verification \\
    \hline

    $d_5$    &    Hardware-based protection \\
    \hline

    $d_6$    &    Secure software stack \\
    \hline

    $d_7$    &    CVE scanning \\
    \hline

    $d_8$    &    Actuator command plausibility checks \\
    \hline

    $d_9$    &    Anti-malware software \\
    \hline

    $d_{10}$    &   Input sanitization  \\
    \hline

    $d_{11}$    &   Redundant inference   \\
    \hline

    $d_{12}$    &   Public key infrastructure  \\
    \hline

    \end{tabular}
\end{table}

\subsection{Mathematical Modeling}    
    The main goal of this work is to construct an attack-defense tree for connected and autonomous vehicles (CAVs), enabling a systematic assessment of system vulnerabilities. Through this representation, it becomes possible to identify and analyze potential security risks that may compromise the functionality, safety, and reliability of CAV systems. By mapping the relationships among different attack vectors and corresponding countermeasures, the attack-defense tree provides a structured way to understand how various threats interact and how they can be mitigated effectively.

    By identifying all possible attack paths, we can focus our defensive efforts on the most vulnerable components of the system. This approach ensures that resources are allocated efficiently, prioritizing the protection of critical assets and communication links within the CAV network. Furthermore, the structured nature of the attack-defense tree facilitates the identification of common attack patterns, helping to develop proactive defense strategies rather than relying solely on reactive measures.
    
    When mathematical modeling is incorporated into the attack-defense tree, it enables quantitative evaluation of the likelihood and impact of different attacks, as well as the effectiveness of existing and proposed defense mechanisms. Such a mathematical framework supports data-driven decision-making, allowing security analysts to prioritize countermeasures based on risk levels and potential system impact. To this end, in the following subsection, we present a mathematical analysis derived from the developed attack-defense tree for CAVs, providing insights into the robustness and resilience of the system against cyber threats.
    
    Mathematically, if the attack leaf vulnerability indices is $\nu(\mathcal{I}_i)$, then, we can write as follows.
    
\begin{equation}
    \nu(\mathcal{I}_i) = \left\{\begin{array}{cl} \text{max}\left\{\alpha \cdot (1 - (n_{d_j}/5)), \alpha \cdot \text{max}\left\{\beta \right\}\right\}, & \alpha > 0 \\
    \text{max}\left\{(1 - (n_{d_j}/5)), \text{max}\left\{\beta \right\}\right\}/3, & \alpha = 0 \end{array}\right.
    \label{eq:vulnerability}
\end{equation}

    Here, $\alpha$, $\beta$ refer to assigned weights in different defensive measures as presented in Table \ref{tab: conditions} and \ref{tab: IDS}, respectively. And, $n_{d_j}$ is the number of defenses under each attack leaf, and the value is considered to be up to $5$ in line with realistic scenarios. Besides, $\mathcal{I}_i$ the attack leaves, and each leaf might have vulnerabilities that are vulnerable to attack, and the vulnerability index runs from $0$ to $1$ values, with $0$ being the most secure, whereas 1 being the most vulnerable.

\begin{table}[ht]
    \footnotesize
    \centering
    \caption{A summary of considered conditions and weight assignments.}
    \label{tab: conditions}
    \begin{tabular}{m{2cm}|m{3.2cm}|m{2cm}}
    
    \hline \hline

    \textbf{Considered Conditions}   &   \textbf{Countermeasures at the Attack Leaves} & \textbf{Weight Assignment} $\alpha$\\
    \hline

    Condition 1    &    No countermeasures implemented at an attack leaf & 1.00 \\
    \hline

    Condition 2     &   At least one countermeasure is there in one attack leaf & 0.50 \\
    \hline

    Condition 3    &    At least three or more than that countermeasures are implemented in order to protect the attack leaf & 0.00 \\
    \hline

    \end{tabular}
\end{table}

\begin{table}[ht]
    \footnotesize
    \centering
    \caption{A summary of the considered conditions for the intrusion detection capabilities.}
    \label{tab: IDS}
    \begin{tabular}{m{1.4cm}|m{4.7cm}|m{1.4cm}}
    
    \hline \hline

    \textbf{Considered Conditions}   &   \textbf{Intrusion Detection Systems (IDS)} & \textbf{Weight Assignment} $\beta$\\
    \hline

    Condition 1    &    IDS-absent configuration \newline Example: The vehicle/network is not equipped with any intrusion detection functionality. & 1.00 \\
    \hline

    Condition 2     &   Minimal IDS \newline Example: The system employs at least one IDS, such as rule-based to detect known patterns or anomaly-based to detect abnormal behaviors and network traffics. & 0.67 \\
    \hline

    Condition 3    &    Standard IDS \newline Example: The configuration represents a robust intrusion detection capability by using both rule and anomaly detection approaches. & 0.33 \\
    \hline

    Condition 4    &    Enhanced IDS \newline Example: The system is secured by a comprehensive capability by combining traditional rule- and anomaly-based systems along with a learning-based improved detection of novel attacks. & 0.00 \\
    \hline

    \end{tabular}
\end{table}

\begin{figure*} [ht]
\centering
\begin{subfigure}[b]{0.32\textwidth}
	\centering
	\includegraphics[width=\textwidth]{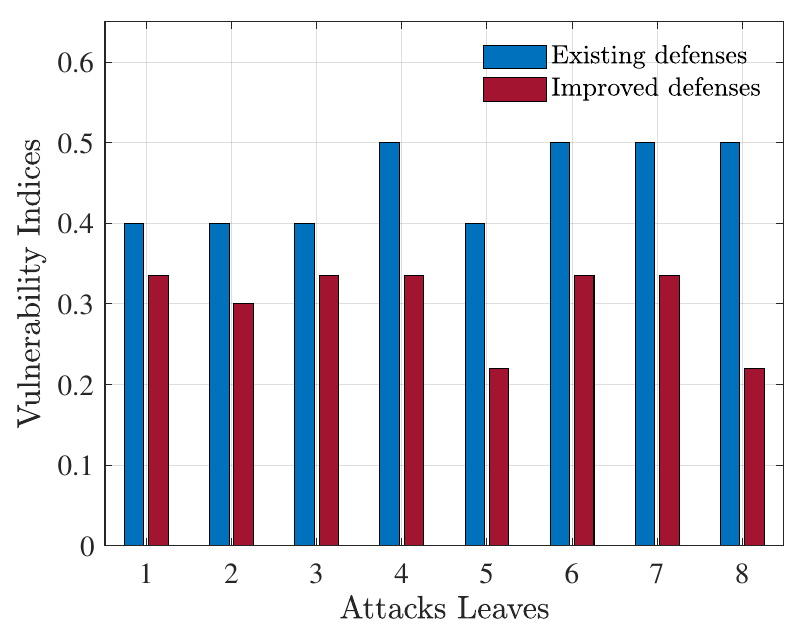}
	\caption{\centering \scriptsize Vulnerabilities at each leaf}
\end{subfigure}
\hspace{.8cm}
\begin{subfigure}[b]{0.32\textwidth}
    \centering
    \includegraphics[width=\textwidth]{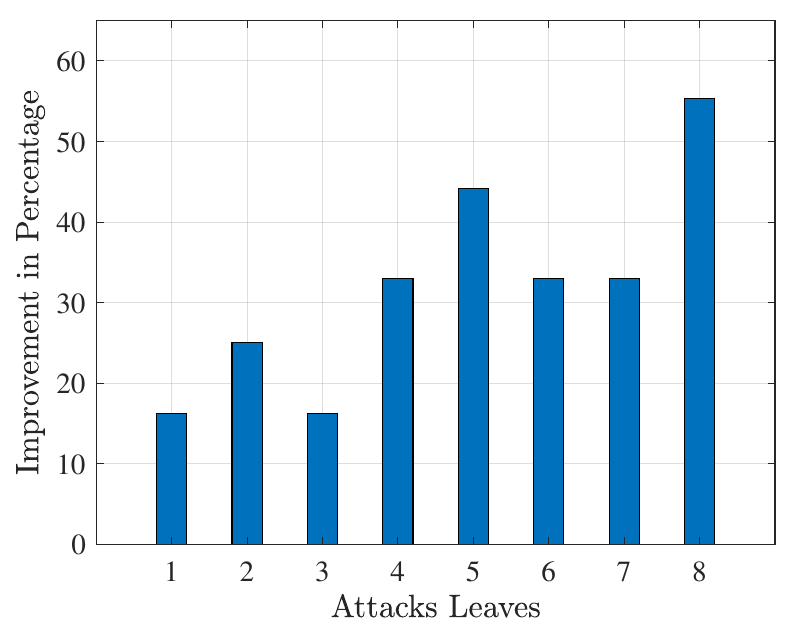}
	\caption{\centering \scriptsize Vulnerability improvement in percentage}
\end{subfigure}
	\caption{Vulnerability evaluating of connected vehicles in quantitative manner.}
	\label{fig: Vulnerability}
\end{figure*}

\subsection{Security Evaluation}
    In Fig. \ref{fig: Attack-Defense}, we presented several example attack and defense scenarios that illustrate potential vulnerabilities in the connected and autonomous vehicle (CAV) system. These examples demonstrate how different attack paths and their corresponding countermeasures interact within the overall system security architecture. Beyond the qualitative representation in the figure, it is also possible to perform a quantitative assessment of the system’s security strength using Equation (\ref{eq:vulnerability}), which provides a mathematical basis for evaluating the level of vulnerability associated with each attack-defense pair.

    To facilitate this evaluation, we list the attack leaves and their corresponding defense mechanisms derived from the attack-defense tree in Table \ref{tab: notations}. This table provides a structured overview of how each attack point in the system is protected and allows for detailed analysis of the security effectiveness of these countermeasures. Furthermore, we revise the defense mechanisms in terms of introducing new countermeasures and enforcing intrusion detection capabilities at each attack leaf. This modification enables a comparative assessment between the current state of defenses and the improved configurations, helping to quantify the impact of enhanced security measures.

    Finally, the results of the evaluated vulnerabilities at each attack leaf are presented as a bar chart in Fig. \ref{fig: Vulnerability}. This visualization provides a clear and concise representation of the relative security strengths before and after implementing the revised defense mechanisms. By comparing these results, we can effectively identify which areas of the system remain most susceptible to attack and where additional reinforcement may be required. Overall, this analysis supports a data-driven approach to improving the resilience of CAVs against evolving cyber threats.

\begin{table*}[ht]
    \centering
    \caption{A summary of attacks, their existing defenses from the attack defense tree, and considered defenses after improvements.}
    \label{tab: notations}
    \begin{tabular}{m{.8cm}|m{3cm}|m{3.2cm}|m{4.3cm}|m{4.4cm}}
    
    \hline \hline
    
    \textbf{Attack Leaves} & \textbf{Descriptions} & \textbf{Existing Countermeasures} & \textbf{Intrusion Detection Systems} & \textbf{After Improvement} \\
    \hline

    1 & In-vehicle CAN bus replay & Authentication tags to verify message integrity and freshness & Rule-based detection of too-frequent, unrealistic, or unexpected sequence ordering messages & Bus segmentation \& gateway filtering \\
    \hline
    
    2 & In-vehicle CAN bus flooding & Rate limiting and message frequency control & Identifying message rate violations that exceed predefined thresholds \newline Detecting sudden deviations from normal traffic patterns & Bus segmentation \& gateway filtering \newline Learning-based capability to recognize unknown attacks \\
    \hline
    
    3 & In-vehicle ECU malicious firmware update & Secure boot and firmware integrity verification & Intrusion detection for update behavior & Hardware-based protection, such as Trusted Execution Environment (TEE) or Hardware Security Module (HSM) \\
    \hline

    4 & Malicious command injection into actuators & Command authentication and cryptographic signatures & IDS-absent configuration & Actuator command plausibility checks \newline Rule-based IDS (e.g., thresholds, rate limits) \\
    \hline

    5 & Rogue roadside unit deployment & Mutual authentication between vehicles and roadside unit & Installed anomaly-based IDS to detect abnormal behaviors and network traffics & Anti-malware software \newline Identity verification with certificates \\
    \hline

    6 & Remote code execution by exploiting software vulnerabilities & Secure software stack & IDS-absent configuration & CVE scanning \newline Anomaly-based IDS for detecting abnormal behaviors and triggering protective actions \\
    \hline

    7 & Certificate misuse or credential theft & Public key infrastructure & IDS-absent configuration & Certificate revocation and monitoring certificate revocation lists \newline Role-based access control \\
    \hline

    8 & Edge AI model poisoning for cooperative perception & Model integrity verification & IDS-absent configuration & Input sanitization \newline Redundant inference from multiple sources \newline Monitoring model update behavior
    \\
    \hline
    
    \end{tabular}
\end{table*}
    
\section{Conclusion}
    As technologies related to connected and autonomous vehicles (CAVs) continue to advance, developing robust cyber security measures becomes essential to protect against potential threats and ensure the safety of passengers and other road users. Attack-defense tree modeling offers a systematic approach to represent a broad range of security threats and serves as a valuable tool for security analysts responsible for safeguarding systems and data. In this paper, we have illustrated potential cyber threats in CAVs and their corresponding countermeasures using attack-defense tree modeling. We then presented a quantitative evaluation of vulnerabilities, highlighting how this approach can facilitate effective security assessments and inform mitigation strategies. For the future, this work can be extended with dynamic dependencies, probabilistic threat propagation, and adaptive countermeasures that can provide a more realistic assessment of vulnerabilities by combining this with tools such as Bayesian networks, Markov decision processes, or game-theoretic modeling to quantify risk, evaluate defense strategies, and enable proactive security measures to protect CAVs against sophisticated attacks.

    \textbf{Acknowledgment:} This research is partly supported by National Science Foundation (NSF) under the grant numbers \# 2010366 and \#~2140729.

\ifCLASSOPTIONcaptionsoff
  \newpage
\fi

\bibliographystyle{IEEEtran}
\bibliography{bibliography.bib}

\end{document}